\documentclass[11pt]{article}
\usepackage{graphicx}
\usepackage{subfigure}
\usepackage{xcolor}
\usepackage{amssymb}
\usepackage{amsthm}
\usepackage{amsmath}
\usepackage{verbatim}
\usepackage{authblk}
\usepackage{lineno}
\usepackage{epstopdf}

\makeatletter   

\newtheorem{theorem}{Theorem}[section]
\newtheorem{definition}[theorem]{Definition}
\newtheorem{proposition}[theorem]{Proposition}
\newtheorem{lemma}[theorem]{Lemma}
\newtheorem{corollary}[theorem]{Corollary}

\newtheorem{assumption}[theorem]{Assumption}

\@addtoreset{equation}{section}

\makeatother 
\def\R{\mathbb{R}}

\def\E{\mathbb{E}}
\def\P{\mathbb{P}}
\def\F{\mathcal{F}}

\def\R{\mathbb{R}}

\def\P{\mathbb{P}}
\def\Y{\mathcal{Y}}
\def\H{\mathcal{H}}
\def\E{\mathbb{E}}

\def\X{\mathbf{X}}
\def\cc{\mathbf{c}}
\newcommand{\one}{\mathbf{1}}
\newcommand{\two}{\mathbf{2}}
\newcommand{\three}{\mathbf{3}}
\def\Cone{\cc^{\one}}
\def\Ctwo{\cc^{\two}}
\def\Ceye{\cc^{\mathbf{i}}}
\def\Xone{\X^{\one}}
\def\Xtwo{\X^{\two}}
\def\Xthree{\X^{\three}}
\def\Xeye{\X^{\mathbf{i}}}
\def\ppi{\boldsymbol{\pi}}
\def\pione{\ppi^{\one}}
\def\pitwo{\ppi^{\two}}
\def\pithree{\ppi^{\three}}
\def\pieye{\ppi^{\mathbf{i}}}

\newcommand{\Cc}{\mathcal{C}}

\newcommand{\Fc}{\mathcal{F}}

\newcommand{\g}{\gamma}

\def \proof{{\noindent \bf Proof}\quad}
\def \ep{\hbox{ }\hfill$\Box$}

\author[1]{O. Mbodji
}
\author[2]{A. Nguyen-Huu}
\author[1]{T. A. Pirvu
}
\affil[1]{{\small McMaster University, 1280 Main Street West, Hamilton, ON L8S 4L8, Canada}}
\affil[2]{{\small CEE-M, University of Montpellier, CNRS, INRA, Montpellier SupAgro, Montpellier, France}}
\renewcommand\Authands{ and }

\title{Optimal Sharing Rule for a Household \\
	with a Portfolio Management Problem\footnote{The authors would like to thank the referees for helpful comments, in particular with situating the present paper in the appropriate literature.
		The usual disclaimer applies.
Adrien Nguyen-Huu is supported by the Chair Energy \& Prosperity, and the Labex Entreprendre. The third author acknowledges the support  of NSERC grant 371653-09 and SSHRC grant 5-26758.}}
\date{\today}

\begin{document}

\maketitle
\begin{center}

%


\mbox{}\\[0pt]

\end{center}
\vspace{-1.4cm}
\begin{abstract}
{We study an intra-household decision process in the Merton financial portfolio problem.
This writes as an optimal consumption-investment problem in finite horizon for the case of two separate consumption streams and 
a shared final wealth, in a linear social welfare setting.
We show that the aggregate problem for multiple agents can be linearly separated in
multiple optimal single agent problems given an optimal sharing rule of the initial endowment.
Consequently, an explicit closed form solution is obtained for each subproblem,
and for the household as a whole.
We show the impact of asymmetric risk aversion and market price of risk
on the sharing rule in a specified setting with mean-reverting price of risk, with numerical illustration.}
\end{abstract}




\section{Introduction}
\label{sec:intro}

{An important dimension of household savings decisions
is the possibility of individual consumption streams out of the common wealth.
In the standard household economics literature, e.g., \cite{browning1996household, browning2000saving},
the decision is most impacted by arbitrage with individual incomes.
It has already been argued in \cite{ortigueira2013important} that the household---as a risk-sharing institution---considers personal savings as contribution to a  group insurance,
and that ignoring intra-household risk sharing introduces a bias in the response of savings to income shocks.
In the present paper, we take a drastically different stand from the literature by focusing
exclusively on the management of savings.
The goal of this paper is more precisely to study the impact of heterogeneous preferences within the household on portfolio initial allocation, in a complex and dynamic financial world.
}

{General intra-household decision problems constitute a well-known challenge.
It is acknowledged that
bargaining concepts (Nash style or Kalay-Smorodinsky, see \cite{manser1980marriage}) are not the
only "collective" decision process alternative to modeling the household as a single decision unit.
In his seminal contribution, Chiappori \cite{chiappori1988rational} minimally defines collective rationality\footnote{For our concern, he defines precisely Collective Rationality of Egoistic Agents (CREA), i.e., a non-cooperative version of CR.}
by simply requiring the household to be on the Pareto-efficient frontier.
This notably leads the household to derive, in his respective labor-consumption problem,
an income sharing rule based on the common initial endowment.
He \cite{chiappori1992collective} then shows in that setting that 
household decisions are efficient if and only if some sharing rule exists.
Obtaining such a rule is our objective, in a very specific context: the Merton portfolio problem.}

{The optimal investment problem is much more involved than a static labor-consumption allocation,
for at first it is dynamical.
Obtaining the initial sharing rule is only the first step of solving the economic problem:
explicit the financial strategy and the optimal consumption path are necessary to fully understand intra-household decisions, in time.
In a dynamic extension of \cite{chiappori1988rational, chiappori1992collective},
Mazzocco \cite{mazzocco2007household} underlines that household decisions must be specified to be
under some commitment or non-commitment paradigm:
if the evolution of members endowments moves away from the sharing rule,
anticipation of that fact must be taken into present negotiation.
Under commitment, members sticks to the sharing rule while without it, the solution involves a game-theoretic approach\footnote{In the continuous time setting,
the non-commitment approach and a relative game-theoretic solution are usually invoked in time-inconsistent optimal control problems, see e.g. \cite{ekeland2012time} in a close setting.
This is actually the case here when agents have heterogeneous preferences. This is not investigated here but should definitely be a topic for further research.}. 
See also \cite{chiappori2017static} for a recent review on that topic.
For the sake of simplicity, we remain under the commitment assumption.
Yet in the easiest setting, the problem remains difficult if the intra-household interactions are not specified.
That is why we set the problem for a linear household Welfare function based on separable Von-Neuman Morgenstern utility preferences.
As we obtain the optimal sharing rule in this setting (see Theorem \ref{theorem} hereafter),
we are inclined to conjecture \emph{ex-post} that some collective rationality arises from it (see in particular \cite{chiappori1988rational}, p. 74).}

{How does the problem writes?
Specifically, the household (e.g. spouses) involves two separated utility functions $U_1$ and $U_2$
for consumption streams of money $c^1$ and $c^2$ respectively,
and two distinct (non necessarily constant nor equal) discount rates $\beta^1$ and $\beta^2$ to measure their individual impatience.
Additionally, since we consider problem in finite horizon $T$,
they share a common utility function $U_3$
to evaluate together the resulting terminal wealth $X_T$, discounted with 
rate $\beta^3$.
This wealth is obtained as the result of a financial portfolio strategy on 
the time interval $[0,T]$,
starting from an initial endowment $X_0=x$ and invested in risky or riskless assets.
We pose the problem of maximizing 
\begin{equation}
\label{eq:problem 1}
\E\left[\int_0^T e^{\left(-\int_0^t \beta^1(s)ds\right)} U_1(c_t^1)dt
+\int_0^T e^{\left(-\int_0^t \beta^2(s)ds\right)}U_2(c_t^2) dt
+e^{\left(-\int_0^T \beta^3(s)ds\right)} U_3(X_T)\right] \;,
\end{equation}
by choosing the optimal consumption rates $(c^i_t)_{t\in [0,T]}$ for $i=1,2$ and the portfolio allocation $(X_t)_{t\in [0,T]}$.
Having now sketched the mathematical problem,
we may comment on it from different standpoints.}

{It is important to understand that
the portfolio is a self-financing portfolio, i.e., 
it starts with a given initial wealth and does not undergo any additional injection of savings.
The fundamental factors influencing the sharing rule will
thus be risk aversion embedded in utility functions,
and impatience in individual discount rates.
Our contribution mainly attempt to provide the sharing rule
regarding those typical financial dimensions,
that is, when members of the household have different levels of risk aversion or impatience.
In this respect, our article  really stands as a contribution to the portfolio management literature.}

But in the later research field, problem \eqref{eq:problem 1} is relatively new.
In the classical portfolio management problem,
the single agent model is the default representation.
It has already been argued \cite{Six_optimalconsumption} that
consumption and terminal wealth should be evaluated distinctly,
as the nature of the reward is different 
(empirical works \cite{meyer2005} highlight a greater risk aversion toward consumption than toward wealth).
In \cite{Six_optimalconsumption}, Six makes that distinction, yet for a single agent.
He shows, as we do, the separability of the problem,
and that the allocation of money to consumption (the \textit{consumption satisfaction proportion}, or CSP) drastically depends---but monotonously---on the initial wealth.
By introducing two separate consumption streams,
we go further than Six \cite{Six_optimalconsumption},
since there is an intricate dependence of each consumption stream with the common resulting wealth.
We show that our model, unlike the one of \cite{Six_optimalconsumption},
can exhibit a hump shaped consumption satisfaction proportion.

What other insights do we learn from this model?
We end the paper with a numerical application with closed form solutions
based on the consumption-savings problem of Wachter \cite{Wachter}.
Especially, we show that the previously mentionned \textit{consumption satisfaction proportion} increases with the initial endowment,
but slower for the more risk averse agent.
When the initial endowment is large enough, the less risk averse agent  allocates relatively more money to future consumption.
This mainly shapes the initial sharing rule, as expected.
This allows us to study consumption satisfaction proportion dependence with respect to the market price of risk and risk aversion.
The numerical results revealed that, unlike the effect of market price of risk change which is marginal, a change in risk aversion
can significantly impact the consumption satisfaction proportion.
Those intuitions, developed in Section \ref{sec:numerics},
are for the benefit of portfolio managers.
As we mentioned it a the very beginning, 
we acknowledge that the study of a household financial portfolio problem is at odd
with existing literature focusing on labor-consumption-leisure arbitrage, and that few household data would be available to help reveal the pertinence of 
the implemented setting.
Nevertheless, 
the application seems much more realistic when one thinks of the situation of a mutual fund portfolio manager working for a pool of heterogeneous clients,
or in the framework of a hedge fund optimal dividend distribution from the shareholders perspective.
If household decision theory is our starting point to introduce the problem,
the applications are more numerous in the financial industry.
We comment shortly in the paper on the possibility to extend the setting to $n>2$ consumption streams.
The reader will easily be able transpose te present results to a much larger pool of agents.

{The remainder of the article is organized as follows.
Section \ref{sec:model} introduces the financial market, assumed to be complete, portfolios and their properties.
Section \ref{sec:main result} develops the central theorem, i.e., the optimal initial sharing rule.
Accordingly, the problem can be separated in three subproblems which are solved explicitely via classical duality techniques, see \cite{KaratzasSmallInvestor}.
We comment also on how \cite{Six_optimalconsumption} extends non trivially from one to two agents, and how to extend the setting to more than two.
Section \ref{sec:numerics} presents the specific case based on power utility functions and a mean reverting price of risk,
first introduced in \cite{Wachter}.}

\section{The financial market and portfolio properties}
\label{sec:model}
\subsection{A complete financial market}
\label{sec:complete market}

We start by describing the dynamics of asset prices.
We consider a classical framework of a complete financial market in continuous time,
with one riskless asset $S^0$ (a bank account) and $d$ risky assets $(S^1, \ldots, S^d)$, i.e. stocks.
It is important to notice that the methodology we developed does not extend naturally to incomplete markets. 
This is because of the non uniqueness of the martingale measure, 
which would forbid to uniquely define the utility price of risks\footnote{The Backward Stochastic Differential Equations approach may work in the incomplete market setting
	within our context, but we leave this as topic of future research. We refer to \cite{HIM} for this approach in the expected (power and exponential) utility maximization of terminal wealth, \cite{HRIH} for a wider class of utility functions and  \cite{CH} for the introduction of consumption streams.}.

The riskless asset will evolve at the interest rate $(r_t)_{t\in[0,T]}$, that is,

{$$
dS^0_t=r_t S^0_t dt\;, \quad 0\le t\le T\;,
$$
with $S^0_0=s_0$.
To model risky assets, we consider a filtered probability space 
$(\Omega, \F, \P)$ 
supporting a standard $d$-dimensional Brownian motion 
$W:=(W_t)_{t\in [0,T]}=(W^1_t,\ldots,W^d_t)^T_{t\in[0,T]}$.
As it is usually assumed, the filtration $(\F_t)_{t\in[0,T]}$ is 
the augmentation under $\P$ of the natural filtration of $W$.
The risky assets then follow a generalized Black-Scholes model (an It\^o diffusion process):
\begin{equation}
\label{eq:diffusion}
dS^i_t=S^i_t\left(b_i(t)dt+\sum_{j=1}^{d}\sigma_{ij}(t)dW_t^j \right)\; , \quad 0\leq t\leq T\;,
\end{equation}
with $(S^1_0,\ldots,S^d_0)=(s_1,\ldots,s_d)\in (0,\infty)^d$.
The vector of mean rates of return 
\newline $b(t):=(b_1(t)\ldots b_d(t))^T_{ t\in [0,T]}$ 
and the diffusion matrix 
$\sigma(t):=(\sigma_{ij}(t))_{1\leq i,j \leq d, t\in [0,T]}$ 
are assumed to be adapted to the filtration $\F$.
Moreover, $\sigma(t)$ is assumed to be invertible for all $t$.
The interest rate process 
$(r(t))_{ t\in  [0,T]}$ can also be made stochastic if it is assumed to be adapted.
We assume that $b$, $\sigma$ and $r$ are such that 
the stochastic differential equation \eqref{eq:diffusion} has a unique strong solution
}

{As usual, we use the riskless asset as a num\'eraire and replace asset prices by their discounted counterpart.
The discount factor is defined by
\begin{equation}\label{Dt}
D_t:=\exp\left(-\int_0^t r(u)du\right)\;, \quad 0\le t\le T\;.
\end{equation}
and for a generic process $Y_t$,
$\tilde{Y}_t:=Y_tD_t$ denotes the discounted version.}
%


{Importantly, we assume that the financial market is complete, in the sense of \cite{harrison1981martingales}.
It will mean that no arbitrage is possible on the market and that utility price of risks can be uniquely determined.
It is defined in our setting by the existence of a unique $\P$-equivalent martingale measure 
$\tilde{\P}$.
If we define the price of risk process $\theta(t):=\sigma(t)^{-1}(b(t)-r(t){\mathbf{1}})$, 
for $t\in [0,T]$,
then market completeness translates into some integrability conditions on $\theta$, so that
$\tilde{W_t}:=W_t + \int_0^t \theta(s) ds$ is a Brownian motion under $\tilde{\P}$, 
see \cite{KaratzasShreve1987}.
We set $\tilde \E$ the expectation operator under $\tilde{\P}$.
We can also specify the Radon Nikodym derivative process of 
$\tilde{\P}$ w.r.t. $\P$ as the process
\begin{equation}\label{Zt}
Z_t :=\exp\left\{-\sum_{i=1}^d \int_0^t \theta_i(s)dW_s^i -\frac{1}{2}\int_0^t ||\theta(s)||^2 ds\right\}\;, 
\quad \text{ for } t\in[0,T] \;.
\end{equation}
A sufficient condition for market completeness to hold is that the process $(Z_t)_{t\in [0,T]}$ is a (true) $\P$-martingale.
We thus assume the latter throughout the paper.
Sufficient conditions for $Z$ to be a martingale are Novikov and Kazamaki conditions \cite{KaratzasShreve1987}. 
If the process $\theta(t), t\in[0,T]$ is Markovian, \cite{WH} provide finer sufficient conditions. Those are not the concern of this paper.}

\subsection{Admissible portfolios}
\label{sec:portfolio}

The above framework has been considered by \cite{KaratzasSmallInvestor} 
for a single investor and \cite{Six_optimalconsumption} focused on the special case of 
an investor with two different power utilities:
one for consumption and one for final wealth.
We consider here two consumption streams and one common terminal portfolio value evaluation.
We comment the generalization to an arbitrary number of consumption streams and terminal values
in subsection \ref{rem:generalization}.
	
{Our first concern is to represent the asset allocation.
	For that purpose we define a \textit{portfolio strategy}
	as a $\F$-adapted $\R^d$-valued process
	$\pi:=\{\pi(t)=(\pi_1(t),\ldots,\pi_d(t))^\top\}$,
	where $\pi_i(\omega) \in L^2([0,T])$ for
	$\P$-almost every $\omega\in \Omega$.
	The interpretation is natural: for $i=1,\ldots,d$, $\pi(t)$ denotes the number of shares of asset $i$ held in the portfolio at time $t$.}
	
{We then introduce \textit{consumption streams} as $\F$-adapted processes $(c_t^1,c_t^2)_{t\in [0,T]}$.
	They are assumed to take non-negative values and be such that $C(\omega):=c^1(\omega)+c^2(\omega)$ 
	is in $L^1([0,T])$ for $\P$-almost every $\omega\in \Omega$.}

{The corresponding \textit{wealth process} $X:=(X_t)_{t\in [0,T]}$ is uniquely defined by
	\begin{equation}
	\label{soldiffX}
	X_t = \frac{1}{D(t)}\left( x+\int_0^t   
	\left( \left(\pi^T(s)b(s)-C_s)\right)D(s)ds + \pi^T(s)\sigma(s)D(s)dW_s \right)\right),
	\end{equation}
	or equivalently by the discounted version
	\begin{equation}
	\label{soldiffXdisc}
	\tilde{X}_t = \left( x+\int_0^t   
	\left( \left(\tilde{\pi}^T(s)b(s)-\tilde{C}_s)\right)ds + \tilde{\pi}^T(s)\sigma(s)dW_s \right)\right)\; .
	\end{equation}
\begin{definition}
		\label{def:admissibility}
		A triplet $(\pi, c^1, c^2)$ 
		of strategy and consumption processes 
		is said to be admissible 
		for the initial endowment $x\geq 0$ 
		if the corresponding wealth process $X_t, t\in[0,T]$ satisfies 
		$X_t\geq 0$ for $[0,T]$ $\P$-a.s.
		We call $A(x)$ the class of 
		admissible processes $(\pi, c^1, c^2)$ 
		for initial wealth $x$, 
		and $P(x)$ the subset of $A(x)$
		composed of triplets of the form $(\pi, 0,0)$.
\end{definition}}

{The set $P(x)$ describes self-financed portfolio strategies in the usual sense.	}

\subsection{Supermartingale properties}
\label{sec:martingale}

{Using standard results, 
	we develop here intutions on how the portfolio and consumption will be managed
if the problem \eqref{eq:problem 1} can be separated in three.
The results below are mostly technical and can be skipped at first reading.}

Admissibility of strategies can be expressed as a no-arbitrage or martingale property.
	Indeed, one can write equation \eqref{soldiffX} under $\tilde \P$:
	\begin{equation}\label{eqXt}
	\tilde{X}_t  +\int_0^t \tilde{C}_s ds=x+\int_0^t \tilde{\pi}^T(s)\sigma(s)d\tilde{W}(s) \; ,
	\end{equation}
	and notice that if the corresponding triplet $(\pi, c^1, c^2)$ is admissible,
	then  the left-hand side of \eqref{eqXt} is non negative, 
	and the right-hand side is a local martingale under $\tilde{\P}$.
	It follows that the left-hand side, 
	and hence also $\tilde{X}_t$, 
	is a non negative super-martingale under $\tilde{\P}$.
	Now, 
	if $\tau_0:=T\wedge\inf \{0\leq t\leq T, X_t=0\}$, 
	then $X_t=0$ for all $t \in [\tau_0,T]$ on $\{\tau_0>-\infty\}$.
	The super martingale property in \eqref{eqXt} yields 
	\begin{equation}\label{ineqXC}
	\tilde{\E}\left[\tilde{X}_T + \int_0^T \tilde{C}_t  dt\right] \leq x \;.
	\end{equation}
	An immediate intuition is that after applying the sharing rule on the initial endowment,
	a portfolio dedicated to consumption of one or the other member of the household,
	being assigned her initial endowment,
	should optimally end with no wealth at all.
	We thus introduce for any $x>0$ the set $C(x)$ as
	the set of consumption streams $c(t)$ such that
	\begin{equation}\label{ineqC}
	\tilde{\E}\left[\int_0^T \tilde{C}_t dt\right] \leq x  \; ;
	\end{equation}
	and $D(x)$ the subset of $C(x)$ when equality holds. By linearity of the expectation,
	$C=c^1+c^2\in C(x_1+x_2)$ if $c^1\in C(x_1)$ and $c^2\in C(x_2)$ (and the same by replacing with the sets $D$).
	The wealth process $X$  corresponding 
	to any $C\in D(x)$ satisfies
	\begin{equation}
	\label{eq:construction of pi}
	X_t=\tilde{\E}\left[\int_t^T \tilde{C}_s ds|\F_t\right],\quad 0\leq t\leq T \; .
	\end{equation}
	In particular, 
	$X_T=0$ $\P$-a.s., as foresaw.
	We implicitely use expression \eqref{eq:construction of pi} to construct a portfolio from a consumption process.
	
Similarly for the final wealth, we define $L(x)$ (resp. $M(x)$) the class of non negative random variables $L$ on $(\Omega,\F_T,\tilde{\P})$ which satisfy
	\begin{equation}\label{ineqXT}
	\tilde{\E}\left[ L  \right]  \leq x \ \left(\text{ resp. }=x \right) \;.
	\end{equation}
	As shown in \cite{Six_optimalconsumption} by applying the martingale representation theorem,
we can construct a portfolio strategy $\pi$ for every given $c^1+c^2\in C(x)$, 
such that $(\pi,c^1,c^2)\in A(x)$.
Again, we can avoid explicating $\pi$ which is deduced from other quantities.
The set $C(x)$ 
consists of exactly those ``reasonable'' consumption processes, 
for which the couple of investors, 
starting out with wealth $x$, 
is able to construct an admissible portfolio.

Reciprocally for every $L\in L(x)$
there exists a trio $(\pi, c^1, c^2)\in A(x)$, which can be explictely constructed (see \cite{Six_optimalconsumption}),
with corresponding wealth process $X$, for which
$X_T=L$ $\P$-a.s.
Specifically
there exists a portfolio strategy $\pi \in P(x)$ 
with corresponding wealth process
\begin{equation}
X_t=\tilde{\E}[L | \F_t], \quad\text{ for\; } t\in [0, T].
\end{equation}
Thus, the extreme elements of $L(x)$ (i.e., in $M(x)$)
are attainable by strategies that mandate zero consumption.
Indeed
according to \eqref{eqXt}, 
the set $P(x)$ corresponds to strategies 
such that $X_T$ belongs to $M(x)$.
From now on it will be more convenient to invoke the wealth process $X$ rather than the precise portfolio strategy $\pi$ that
needs to be implemented to obtain the former.
	
In the admissible case, $X_T\geq 0$ and $C_t\geq 0$
for all $t\in [0,T]$.
Thus,
$(\pi,c^1,c^2)\in A(x)$
implies
$c^1,+c^2\in C(x)$,
and $X_T\in L(x)$ implies 
inequality condition \eqref{ineqC}
which, together with \eqref{ineqXT},  
turns out to be also sufficient for admissibility.

\section{The household problem}
\label{sec:main result}

\subsection{Risk aversion, impatience and additional elements}
\label{sec:specification of agents}

Each member of the household is endowed with a utility function $U_i$ for $i=1,2$
	applied to consumption streams,
	and they share a common utility $U_3$ for the final wealth.
	All three functions satisfy the following assumption.
	\begin{assumption}
		\label{AssUtility}
		The function $x\mapsto U(x)$ is a strictly increasing, 
		strictly concave real-valued function 
		in $C^2([0,\infty])$ such that 
		$x\mapsto U''(x)$ is non decreasing, 
		$U(0)\geq -\infty$ and 
		$U'(\infty)=0$. 
	\end{assumption}
	Assumption \ref{AssUtility} designates a class of functions which includes CARA and HARA utility functions.
	Let us drop the $i$ notation for the moment, as what follows stands for the three functions under Assumption \ref{AssUtility}.

{Under Assumption \ref{AssUtility}, a marginal utility $U'$ is defined from 
	$[0,\infty)$ onto $[0,U(0)]$.
	We denote $I:= (U')^{-1},$ the inverse functions of 
	the marginal utility (later indexed with $i=1, 2, 3$).
	Note that we allow for $U(0)=-\infty$ or $U'(0)=\infty$.
	Because 
	$U':[0,\infty]\rightarrow [0,U'(0)]$ 
	is strictly decreasing, 
	it has a strictly decreasing inverse 
	$I:[0,U'(0)]\rightarrow [0,\infty]$. 
	We extend $I$ to be a continuous function on the entirety of $[0,\infty]$ 
	by setting $I(y)=0$ for $U'(0)\leq y\leq \infty$, 
	and note that 
	\begin{equation}\label{ineqU}
	U(I(y))\geq U(c)+y I(y) -yc, \; \text{for}\; (y,c)\in (0,\infty)\times [0,\infty)\;.
	\end{equation}
	This equation will turn to be essential in order to apply duality results in the flavor or \cite{KaratzasSmallInvestor}.}

To each utility valuation corresponds a discount rate $\beta^i$
	assumed to be $\F$-adapted and bounded for all $t$ uniformly $\P$-almost surely.
	An important triplet of processes for later are
	the state price processes $\zeta^i$ which depend on the discount rates for $i=1,2,3$:
	$$
	\zeta_t^i:=Z_tD_t\exp\left(\int_0^t \beta^i(u) du\right)=\frac{\tilde{Z}_t }{B_t^i} \;.
	$$
	where the process $Z_t, t\in[0,T]$ is the Radon Nikodym process given by \eqref{Zt},
	$D_t$ the market discount factor from \eqref{Dt},
	and
	\begin{equation}\label{Bi}
	B^i_t:=\exp\left(-\int_0^t \beta^i(s)ds\right), \quad i=1,2,3
	\end{equation}
	are notations for the discount factors for their respective utility functions.
	Recall now the general discount factor process $(D_t)_{t\in [0,T]}$,
	for it is used in the definition of the following functions defined on $\R_+$:
	\begin{equation}\label{eu}
	\H_i(y) := \tilde{\E}\left[ \int_0^T D_u I_i(y\zeta_u^i)du|\F_t\right]  \; \text{for }\; i=1, 2 \;,
	\end{equation}
	and 
	\begin{equation}\label{H3t}
	\H_3(y) := \tilde{\E} \left[ D_T I_3(y\zeta_T^3)|\F_t\right] \; .
	\end{equation}
	\begin{assumption}
		\label{AssumHi}
		The functions $\H_i$ are finite, i.e. for all $y\in (0, \infty)$ and $i=1,2,3$, $\H_i(y)<\infty$.
	\end{assumption}
	The $\H_i$ functions basically stand for the dual functions, which associate a monetary worth to a utility level $y$, discounted and expected properly,
	and play a determinant role in fixing the sharing rule below. 
	\begin{lemma}
		\label{lem:karatzas}
		For $i=1, 2, 3$, 
		$\H_i$ is a continuous function, 
		strictly decreasing on $(0,\infty)$ with
		$\H_i(0)=\infty$ and $ \H_i(\infty)=0$
	\end{lemma}
	
	The above Lemma is proved in \cite{KaratzasSmallInvestor}.
	Since the above functions are strictly monotonous as well and we can define $\Y_i:=\H_i^{-1}$ the inverse of the function $\H_i$ for $i=1,2,3.$ 
	Let us notice that 
	$\Y_i: [0,\infty] \rightarrow [0,\infty]$.
	We now have properly introduced the dual elements for each problem. 
	We end this section by introducing a technical assumption about them.
	
	\begin{assumption}
		\label{AssumUbound}
		For all $y\in (0, \infty)$,
		\begin{eqnarray*}
			\E\left[\int_0^T B_t^1|U_1(I_1(y\zeta_t^1))|dt+\int_0^T B_t^2|U_2(I_2(y\zeta_t^2))|dt
			+ B_T^3|U_3(I_3(y\zeta_T^3))|\right] < \infty \;.
		\end{eqnarray*}
		Furthermore assume that the functions
		$$y\mapsto \E\left[\int_0^T B_t^i|U_i(I_i(y\zeta_t^1))|dt\right],\,\,\, i=1,2,\quad y\mapsto   \E B_T^3|U_3(I_3(y\zeta_T^3))|, $$
		are differentiable and the derivatives with respect to $y$ can be taken under the expectation and integral signs.
	\end{assumption}
	For example, this assumption translates into the finitness of nth order moment of the process $Z$
	defined in \eqref{Zt}, when utilities are of the nth power type.
	All the above assumptions are therefore purely technical and are attributable to agents' preferences.
	
\subsection{Value functions}
\label{sec:value functions}

{We now formalize the main problem \eqref{eq:problem 1}.
According to definitions \eqref{Bi},
we can now rewrite \eqref{eq:problem 1} as the objective function

\begin{equation*}
\label{J}
J(x;\pi,c^1,c^2):=E\left[\int_0^T (B^1_tU_1(c_t^1)+B^2_tU_2(c_t^2))dt +B^3_TU_3(X_T)\right].
\end{equation*}
To our optimization problem, and for 
a given initial wealth $x>0$ shared by the agents,
we can associate the value function 
\begin{equation}
\label{V}
V(x):=\hspace{-0.4cm} \sup\limits_{(\pi,c^1,c^2)\in \tilde{A}(x)}   \hspace{-0.4cm}J(x;\pi,c^1,c^2)\;,
\end{equation}
where the set  $\tilde{A}(x)$ is a subset of $A(x)$ defined simply to ensure that the expectation, that is

$$
\tilde{A}(x):=\left\{(\pi,c^1,c^2)\in A(x): |J(x;\pi,c^1,c^2)| < \infty \right\}.
$$
Notice that for any $x>0$, $\tilde{A}(x)=A(x)$ as soon as $U_i(0)>-\infty$ for $i=1, 2, 3$.
From now on we will exclusively focus on admissible triplets in $\tilde{A}(x)$.}

For a twice differentiable function $V$, we recall that the relative risk aversion is defined by
	\begin{equation}
	\label{eq:RRA}
	x\mapsto R(x):=-\frac{xV''(x)}{V'(x)} \; .
	\end{equation}

Anticipating theorem \ref{theorem} below---that an optimal sharing rule exists---we
define the value functions associated to the three sub-problems relative to the three respective utilities.
For a given $x$, which is now the personal endowment following from the sharing rule,
let $V_1$ be the value function 
\begin{equation}
\label{V_1}
V_1(x):= \sup \limits_{(\pi^1, c^1)\in A_1(x)} J_1(x;\pi^1,c^1) = 
 \sup \limits_{(\pi^1, c^1)\in A_1(x)}  E\left[\int_0^T B^1_tU_1(c_t^1)dt\right]
\end{equation}
with $
A_{1}(x):=\left\{ (\pi^1, c^1)  :  (\pi^1, c^1, 0)\in A(x) \text{ and }|J_1(x;\pi^1, c^1)| < \infty \right\} \;.
$
The value function $V_2$ and the set $A_2(x)$ for all $x>0$ are defined in a strictly similar fashion.
For the terminal wealth, we set 
\begin{equation}
\label{V_3}
V_3(x):= \sup \limits_{\pi^3\in P(x)} J_3(x;\pi^3) = 
\sup \limits_{\pi^3\in P(x)}  E\left[\int_0^T B^3_tU_3(X_T)dt\right]
\end{equation}
Notice that we conveniently use subsets of $A(x)$ ($A_1(x), A_2(x) $ and $P(x)$) to facilitate aggregation.

\subsection{Extensions}
\label{rem:generalization}

Our model can be easily extends to more than two consuming agents.
The extension to more than one evaluation of the terminal wealth is
however a matter of definition.
The function $J$ is indeed defined with only one portfolio strategy
and one terminal value: the linearity of $X_t$ in the financial strategy $\pi_t$
allows to separate into three sub-problems as asserted by Theorem \ref{theorem}.
However, changing the  third term in the objective function $J$
for a term like
$$
\E \left[e^{\left(-\int_0^T \beta^1 (s)ds\right)}U_4 (X^1_T)+ e^{\left(-\int_0^T \beta^2 (s)ds\right)}U_5(X^2_T)\right] \;,
$$
implies to define how agents $1$ and $2$ proceed.
If they share an initial wealth, a common portfolio, and decide at $T$ to split the final wealth,
then the problem is strictly equivalent to \eqref{V} with $\beta^3=0$ and
$$
U_{3}(x) := \sup_{\epsilon \in [0,1]} e^{\left(-\int_0^T \beta^4 (s)ds\right)}U_4 (\epsilon x)
+ e^{\left(-\int_0^T \beta^5 (s)ds\right)}U_5 ((1-\epsilon) x)\;.
$$
If one wants to distinguish agents portfolios, then he shall redefine $J$
in order to separate trading portfolios and consumption portfolios from $t=0$.
The problem can be solved by using the value function approach of Theorem \ref{theorem}.

The utilities of the two agents consumption and the utility of the final wealth can be aggregated by Pareto weights. 
These weights are a reflection of the bargaining power of the two agents. In this case, for a given $x>0$, we define the value function $V$ evaluated at $x,$ i.e., $V(x)$ by
\begin{equation}
\label{Vpareto}
V(x):=\sup \left\{J(x;\pi,c^1,c^2):(\pi,c^1,c^2)\in \tilde{A}(x)\right\}\;,
\end{equation}
where
\begin{equation*}
J(x;\pi,c^1,c^2):=E\left[\int_0^T \mu^{1}(t)B^1_tU_1(c_t^1)+\mu^{2}(t)B^2_tU_2(c_t^2))dt +(1-\mu^{1}(t)-\mu^{2}(t))B^3_TU_3(X_T)\right].
\end{equation*}  
Here $\mu^{1}(t), \mu^{2}(t)$ are Pareto weights in $[0,1]$ and $t\in[0,T].$ Equation above can be rewritten as

\begin{equation*}
J(x;\pi,c^1,c^2):=E\left[\int_0^T \bar{B}^1_tU_1(c_t^1)+\bar{B}^2_tU_2(c_t^2))dt + \bar{B}^3_TU_3(X_T)\right],
\end{equation*}
with 
\begin{equation}\label{Biii}
\bar{B}^i_t:=\exp\left(-\int_0^t \bar{\beta}^i(s)ds\right),\quad i=1,2,3,
\end{equation}
and $\bar{\beta}^i(t),\, t\in[0,T],\, i=1,2,3$ appropriately chosen.

\section{Solution to the problem}

We start by solving problem \eqref{V_1} in subsection \ref{subsec:solve consumption},
and follow with problem \eqref{V_3} in subsection \ref{subsec:solve wealth}.
The optimal sharing rule and the relation with problem \eqref{V} are then presented in subsection \ref{sec:sharing rule},
problem \eqref{V} being solved at this moment.

\subsection{Solving optimal consumption problems}
\label{subsec:solve consumption}

We hereby solve the problem \eqref{V_1} by naturally extending the solution to the other household member.
The setting and the assumptions especially allow to elicitate the explicit optimal consumption path.

\begin{proposition}
	\label{PropCons}
	Let $x_1\ge 0$. Then
	$
	V_1(x_1)= J_1(x_1;\pione(x_1),\Cone(x_1))$ 
	where for all $t\in [0,T]$,
	\begin{equation}
		\label{coptimal1}
		\Cone_t(x_1) := I_1(\Y_1(x_1)\zeta_t^1)\;,\\
	\end{equation}
	and $\pione(x_1)$ is deduced from the wealth process
	\begin{equation}
		\Xone_t(x_1):= \tilde{\E}\left[\int_t^T  \Cone_s(x_1)D_s ds|\F_t \right]\;.\\
	\end{equation}
\end{proposition}

\proof
We take 
$$
c^1=\hat{c}_t^1:=\frac{x_1}{\tilde{\E}\int_0^T D_t dt} \in D(x_1) \;,
$$ 
so that 
$$
\E\left[\int_0^T B_t^1 U_1(\hat{c}_t^1)dt\right]
=U_1(\hat{c}^1)\E\left[\int_0^T B_t^1 dt\right]<\infty\;.
$$
Notice that $\hat{c}^1\in D(x_1)$ and since
$\tilde{\E}\left[\int_0^T D_t\Ceye_t dt\right] = \H_1(\Y_1(x_1))=x_1$,
$\Ceye_t \in D(x_1)$.  
Inequality \eqref{ineqU} implies that 
for any $c^1 \in C(x_1)$ and $t\in [0, T]$, 
$$
U_1(\Ceye_t)\geq U_1(c_t^1)+\Y_1(x_1)\zeta_t^1 \Ceye_t-\Y_1(x_1)\zeta_t^1 c_t^1\;, \; \P- \text{a.s.}
$$
Therefore, 
\begin{eqnarray*}
	\label{negpart}
	\E\left[\int_0^T B_t^1 U_1^{-}(\Cone_t)dt\right]
	&\leq& \E\left[\int_0^T B_t^1 \left(U_1(\hat{c}_t^1)+\Y_1(x_1)\zeta_t^1 \Cone_t-\Y_1(x_1)\zeta_t^1 \hat{c}_t^1\right)^{-}dt\right]\\
	&\leq& \E\left[\int_0^T B_t^1 \left(U_1^{-}(\hat{c}^1) + \Y_1(x_1)\zeta_t^1 \hat{c}_t^1\right)\right] <\infty \; .
\end{eqnarray*}
Consider the measure on $[0,T]\times \Omega$ 
defined by $d\nu^1(t,\omega)=B_t^1 dt\P(d\omega)$ . 
For any other consumption process $c^1 \in D(x_1)$, 
we have
\begin{equation*}
\int\!\!\int_{[0,T]\times \Omega}U_1(\Cone_t)d\nu^1 
\geq \int\!\!\int_{[0,T]\times \Omega}U_1(c_t^1)d\nu^1 
+ \int\!\!\int_{[0,T]\times \Omega}\Y_1(x_1)\zeta_t^1 \Cone_t d\nu^1 
- \int\!\!\int_{[0,T]\times \Omega}\Y_1(x_1)\zeta_t^1 c_t^1 d\nu^1
\end{equation*}
By using the fact that $\Cone, c^1 \in D(x_1)$,
\begin{equation*}
\E\left[\int_0^T B_t^1 U_1(\Cone_t)dt\right] \geq \E\left[\int_0^T B_t^1 U_1(c_t^{1})dt\right], \quad i=1,2 \;. 
\end{equation*}
This induces optimality of $\Cone$.
\ep

\subsection{Solving the terminal wealth problem}
\label{subsec:solve wealth}

We now turn to the final wealth valuation.
As mentioned in Section \ref{sec:martingale},
we exclusively focus on self-financed portfolios in $P(x)$.

\begin{proposition}
	\label{PropFinal}
	Let $x_3\in \R_+$. Then $V_3(x_3)=J_3(x_3; \pithree(x_3))$ where
	the corresponding wealth process is given by
	\begin{equation}\label{XT3}
	\Xthree_t(x_3):= \tilde{\E}\left[I_3(\Y_3(x_3)\zeta_T^3)D_T|\F_t \right]
	\end{equation}
	with $\Xthree_T(x_3)=I_3(\Y_3(x_3)\zeta_T^3)D_T \in M(x_3)$.
\end{proposition}

\proof
{\bf 1.}
First,
we show that the (implicit) strategy
$\pithree(x_3)\in P(x_3)$
and that the generated portfolio process
$\Xthree(x_3)$ belongs to $M(x_3)$.
According to \eqref{XT3},
we have 
$$
\tilde{\E}\left[ \Xthree_T(x_3) \right]
=\tilde{\E}\left[ I_3(\Y_3(x_3)\zeta_T )\right]=\H_3(\Y_3(x_3))=x_3 \; .
$$
Considering the constant final wealth 
$b:=x_3/ \tilde{\E}[D_T] \in D(x_3)$, 
we get
$$
U_3(\Xthree_T(x_3))\geq U_3(b)+\Y_3(x_3)\zeta_T^3 \Xthree_T(x_3) - \Y_3(x_3)\zeta_T^3 b\;\; \P-a.s.
$$
Therefore, 
$$
\E\left[ B_T^3 U_3^{-}(\Xthree_T(x_3))\right]\leq \E\left[B_T^3 \left( U_3^{-}(b)+\Y_3(x_3)\zeta_T^3 b \right)\right] <\infty\; .
$$

{\bf 2.}
Let's show that 
the optimal strategy requires zero consumption. 
Let $(\pi,c^1,c^2)\in A(x_3)$ with wealth process $X$ be given. 
Define the random variable
$$
B:=\left\{
\begin{array}{lcl}
\frac{x_3}{\tilde{\E}[D_TX_T]}X_T & \text{if} &  \tilde{\E}[X_T]>0\\
b &\text{if}& \tilde{\E}[X_T]=0
\end{array}
\right. \; .
$$
Since $X_T\in A(x_3)$,
$\tilde{\E}\left[D_TX_T \right]\leq x_3$.
Then
$B\in M(x_3)$ and $B\geq X_T$ $\P$-a.s. 
From Section \ref{sec:martingale}, 
there exists a portfolio $\hat{\pi}\in P(x_3)$
with corresponding terminal wealth $\hat{X}_T=B\geq X_T$ $\P$-a.s. 

{\bf 3.}
To reach $V_3$ with this strategy ,
it suffices to proceed as in Proposition \ref{PropCons}:
\begin{equation*}
\E\left[ B_T^3 U_3(\Xthree_T(x_3))\right]\geq 
\E\left[ B_T^3 \left(U_3(\X_T)+\Y_3(x_3)\zeta_T^3(\Xthree_T(x_3) - X_T)\right)\right]
\geq \E\left[ B_T^3 U_3(\X_T)\right].
\end{equation*}
This implies optimality of the portfolio process $\Xthree(x_3)$.
\ep
\subsection{The optimal sharing rule}
\label{sec:sharing rule}

Our main result is twofold.
First, we assert that the aggregate problem can be divided in sub-problems.
Notice that a triplet
of financial strategies $(\pione(x_1)$, $\pitwo(x_2)$, $\pithree(x_3))$
corresponds to the household aggregate strategy,
Each corresponds to a wealth process, see Section \ref{sec:martingale} above.
Linearity makes it possible to define the total portfolio $\ppi$ 
and the aggregate wealth process $\X$ by
\begin{eqnarray}
\ppi(x_1,x_2,x_3):=\pione(x_1)+\pitwo(x_2)+\pithree(x_3) \;, \\
\X(x_1,x_2,x_3):=\Xone(x_1)+\Xtwo(x_2)+\Xthree(x_3) \; .
\end{eqnarray}

\begin{proposition}
	\label{prop:part1}
	For $x\geq 0$, 
	\begin{equation}\label{eqV*}
	V(x)=\max \left\{ V_{1}(a_1)+V_2(a_2)+V_3(a_3) | a_1, a_2, a_3\in[0,\infty); a_1+a_2+a_3=x \right\}\; .
	\end{equation}
\end{proposition}

\proof
For $x\geq 0$, 
we are given an arbitrary triplet $(\pi,c^1,c^2)\in \tilde{A}(x)$ 
with corresponding wealth process $X_t$.
Recall that
$$
\begin{array}{ccc}
a_1:=\tilde{\E} \left[\int_0^T B_t^1 c_t^1 dt\right]\;, & 
a_2:=\tilde{\E} \left[\int_0^T B_t^2 c_t^2 dt\right]\;  \text{ and } &  a_3:= \tilde{\E} \left[B_T^3 X_T\right] \;.\\  
\end{array}
$$
By the super martingale property, $a:=a_1+a_2+a_3\leq x$ and 
by Propositions \ref{PropCons} and \ref{PropFinal},
$$\left\{
\begin{array}{l}
\E\left[B_t^1 U_1(c_t^1)dt \right]\leq J_1(a_1;\pione(a_1), \Cone(a_1))=V_1(a_1), \\
\E\left[B_t^2 U_2(c_t^2)dt \right]\leq J_2(a_2;\pitwo(a_2),\Ctwo(a_2))=V_2(a_2), \\
\E\left[B_T^3 U_3(X_T) \right]\leq J_3(a_3;\pithree(a_3))=V_3(a_3).
\end{array}
\right.
$$
Adding the three terms, we get
\begin{equation*}
V_1(a_1)+V_2(a_2)+V_3(a_3)=J(a; \ppi,\Cone(a_1),\Ctwo(a_2)) \geq  J(a;\pi,c^1,c^2) \; .
\end{equation*}
Taking the supremum over $a_1+a_2+a_3\leq x$ 
and over $(\pi, c^1, c^2)\in \tilde{A}(x)$, 
we get  
\begin{eqnarray*}
	V(x)&\leq& \sup\left\{V_{1}(a_1)+V_2(a_2)+V_3(a_3) : a_1, a_2, a_3\in[0,\infty); a_1+a_2+a_3\leq x \right\}\\
	&=& \sup\left\{V_{1}(a_1)+V_2(a_2)+V_3(a_3) : a_1, a_2, a_3\in[0,\infty); a_1+a_2+a_3= x \right\}
\end{eqnarray*}
from the non-decreasing characteristic of $V_i$ for $i=1,2,3$.
Furthermore, 
by continuity of the function $(a_1,a_2,a_3)\mapsto V_1(a_1)+V_2(a_2)+V_3(a_3)$, 
the supremum above is attained at a point $(x_1,x_2,x_3)$  and
\begin{equation*}
V(x)\leq V_1(x_1)+V_2(x_2)+V_3(x_3) = J(x; \ppi(x_1,x_2,x_3), \Cone(x_1), \Ctwo(x_2)) \leq  V(x)\; .
\end{equation*}
The processes
$\Xone$, $\Xtwo$ and $\Xthree$ are nonnegative, 
so $\X$ is nonnegative:
$\X$ is clearly in $\tilde{A}(x)$.
\ep

The second part of the result is the finding of the optimal sharing rule $x_1+x_2+x_3=x$.

\begin{theorem}
	\label{theorem}
	Let $x>0$.
	Then, $V(x)=V_1(x_1)+V_2(x_2)+V_3(x_3)$
	with initial allocation $x_i$ is given for $i=1,2,3$ by
	\begin{equation}\label{xi}
	x_i = \H_i(\Y(x)) \;,
	\end{equation}
	where $\Y: x\in [0,\infty] \mapsto \Y(x):=\H^{-1}(x):=(\H_1+\H_2+\H_3)^{-1}(x)  \in[0,\infty].$
\end{theorem}

It stems from Proposition \ref{prop:part1}
by saying that the $x_i$ are found by using the envelope theorem,
together with Lemma \ref{LemmaGi} below, which implies from Assumption \ref{AssumUbound} that

$$
V_1'(x_1)=V_2'(x_2)=V_3'(x_3)=\Y_1(x_1)=\Y_2(x_2)=\Y_3(x_3)=y \; ,
$$
i.e., 
$x_i=\H_i(y)=\H_i(\Y(x))$. 

\begin{lemma}\label{LemmaGi}
	For $y>0$, define
	\begin{eqnarray}
	G_1(y)&:=&\E\left[\int_0^T B_t^1 U_1(I_1(y\zeta_t^1))dt\right]\;,\\
	G_2(y)&:=&\E\left[\int_0^TB_t^2 U_2(I_2(y\zeta_t^2))dt\right]\;,\\
	G_3(y)&:=&\E \left[B_T^3 U_3(I_3(y\zeta_T^3))\right]\;.
	\end{eqnarray}
	Then
	\begin{equation}
	G_i'(y)=y\H_i'(y)\quad i=1,2,3
	\end{equation}
	and $x\mapsto V_i(x)\in C^2((0, \infty))$ with
	\begin{equation}
	\label{eq:V'_i}
	V_i'(x)=\Y_i(x)\quad i=1,2,3,\; x\geq 0 \; .
	\end{equation}
\end{lemma}

\proof
According to Assumption \ref{AssumUbound},
we can take derivatives under the expectation and integral signs
to obtain
\begin{eqnarray*}
	G_1'(y)&=&\E\int_0^T B_t^1\zeta_t^1I_1'(y\zeta_t^1)U_1'(I_1(y\zeta_t^1))dt
	=\E\int_0^T B_t^1\zeta_t^1y\zeta_t^1I_1'(y\zeta_t^1)dt \\
	&=&\tilde{\E}\int_0^T y\zeta_t^1I_1'(y\zeta_t^1)dt =y\H_1'(y) \; .
\end{eqnarray*}
Therefore, 
\begin{equation*}
V_1'(x)=\frac{d}{dx}G_1(\Y_1(x))=Y_1'(x)G_1'(\Y_1(x))=\Y_1'(x)\Y_1(x)\H_1'(\Y_1(x))=\Y_1(x) \;.
\end{equation*}
The other derivatives are computed in the same manner.
\ep


Let us give some intuition for this result.  It says that the couple value function
is the optimal aggregation of individual value functions. This is a Pareto allocation type result (see \cite{HuangLitzenberger} for more on Pareto allocation); the novelty is that the Pareto weights are the initial wealth allocation. 

We will also discuss the quantitative consequences of 
the agents specification on the splitting of the initial wealth
suggested by Theorem \ref{theorem} in Section \ref{sec:numerics}.





\subsection{Delineation against the single agent case}
\label{Del}

In this section we assume that the three agents share a common initial 
wealth $x$ and have CRRA type utilities

$$
U_i(x)=\frac{x^{\g_i}}{\g_i} \quad \text{ for }i=1,2,3.
$$ 
Here $1-\g_i\in(0,\infty)$  is the risk aversion of agent $i$.
Notice that $x\mapsto U_i(x),\,i=1,2,3$ satisfy Assumptions \ref{AssUtility}, and
$I_i(x)=x^{\frac{1}{\g_i-1}}.$ As it is foreseeable and proved above in subsection \ref{sec:martingale},
the wealth attributed to consuming agents is integrally consumed by time $T$.
Following \cite{Six_optimalconsumption},
we can define the \textit{consumption satisfaction proportion} (CSP) by $(x_1+x_2)/x$,
where $x>0$ represents the total initial wealth and $x_1, x_2$ are given by \eqref{xi}.

In the one agent case the function $x\mapsto\mbox{CSP}(x)$
is either increasing or decreasing. Let us prove this claim. Recall that

$$
x_1= k_1 y^{\frac{1}{\g_1-1}}, \quad x_2= k_2 y^{\frac{1}{\g_2-1}}, \quad  x=x_1+x_2,
$$
for some positive constants $ k_1,  k_2.$ Thus

$$
\mbox{CSP}(x)= \frac{k_1}{k_2 y^{\left(\frac{1}{\g_2-1}-\frac{1}{\g_1-1} \right)} +k_1}.
$$
Moreover, $x= k_1 y^{\frac{1}{\g_1-1}}+ k_2 y^{\frac{1}{\g_2-1}},$ whence $y=f(x)$ for some
decreasing function $x\mapsto f(x).$ This proves the claim. Thus, if we observe a CSP
function which is not monotone then we infer that it can not be the outcome of a one agent model.
In our two agent model we can observe a CSP function which is hump shaped (not monotone).  
Indeed, assume that $\g_1<\g_3<\g_2.$ Recall in this case

$$
x_1= k_1 y^{\frac{1}{\g_1-1}}, \quad x_2= k_2 y^{\frac{1}{\g_2-1}}, \quad  x_3= k_3 y^{\frac{1}{\g_2-1}}, \quad  x=x_1+x_2+x_3,
$$
for some positive constants $ k_1,  k_2, k_3.$ Since $x= k_1 y^{\frac{1}{\g_1-1}}+ k_2 y^{\frac{1}{\g_2-1}}+k_3 y^{\frac{1}{\g_3-1}},$ then $y=g(x)$ for some
decreasing function $x\mapsto g(x).$ Moreover

$$
1-\mbox{CSP}(x)= \frac{k_3}{k_1 y^{\left(\frac{1}{\g_1-1}-\frac{1}{\g_3-1} \right)}+k_2 y^{\left(\frac{1}{\g_2-1}-\frac{1}{\g_3-1} \right)} +k_3}.
$$
Since the function $y\mapsto k_1 y^{\left(\frac{1}{\g_1-1}-\frac{1}{\g_3-1} \right)}+k_2 y^{\left(\frac{1}{\g_2-1}-\frac{1}{\g_3-1} \right)} +k_3 $ is hump shaped
and the function $x\mapsto g(x)$ is decreasing then CSP function is hump shaped as well. 

Therefore the one agent model and our model are not observational equivalent.


\section{Application}
\label{meanreverting}
\label{sec:numerics}
\subsection{CRRA utilities and mean reverting market price of risk}

We provide here an explicit model of the previously studied framework. 
The three agents share a common initial 
wealth $x$ and have CRRA type utilities

$$
U_i(x)=\frac{x^{\g_i}}{\g_i} \quad \text{ for }i=1,2,3.
$$ 
Each agent has his own constant discount rate $\rho_i$. Next take $d=1$ as in \cite{Wachter} (the extension to multiple stocks is straightforward).
The asset price follows a geometric Brownian motion.
In order to isolate the effects
of time variation on expected returns, 
the risk-free rate is assumed to be constant
and equal to $r\geq 0$ but this assumption can be relaxed.
We fix the volatility $\sigma:=\sigma_{11}\in(0,\infty)$ for \eqref{eq:diffusion}, but
we do not specify the drift $b_1\in\R$.
Instead, we model the price of risk $\theta$ by 

$$
 d\theta_t=-\lambda_{\theta}(\theta_t-\bar{\theta})dt-\sigma_{\theta} dW_t \;, \quad t\ge 0 \; ,
$$
where $(\lambda_{\theta},\sigma_{\theta},\bar{\theta})\in(0,\infty)^3$.
We assume $W=W^{1}$, so that the stock price $S^1_t$ and
the state variable $\theta_t$ are perfectly negatively correlated.
These assumptions are
like those in \cite{Kim}, 
except that the latter allows for imperfect
correlation, and thus incomplete markets. The extension of our results to incomplete markets is a non-trivial issue
and is left as a topic of future research.

The body of academic literature on long
term mean reversion is more tractable than that
on short
term mean reversion. 
A comprehensive study on the existence of 
mean reversion in Equity Prices has been done in \cite{osfc}.  
The primary case for the existence 
of long term mean reversion
was made in two papers published in 1988, 
one by \cite{Summers}, the other by \cite{Fama1}. 
In summary, 
these papers conclude that for period lengths
between 3 and 5 years, 
long term mean reversion was present 
in stock market returns
between 1926 and 1985. 

\subsection{Semi-explicit solutions}

In this framework,
the modeling assumptions of Section \ref{sec:model} are satisfied.
We now seek for explicit formulations in Theorem \ref{theorem}:
we aim at providing the initial
repartition $x_1, x_2, x_3$ such that $x_1+x_2+x_3=x.$ 
We start by defining certain functions and provide the formulation of wealth processes for each agent.
\begin{definition}\label{formulas_AiHi}
Define the functions $s\mapsto A_{ji}(s), j=1,2,3$ 
verifying
\begin{equation}
\label{eq:boundary condition of A}
 A_{1i}(0)=A_{2i}(0)=A_{3i}(0)=0\;, \quad  0\le s \le T\; ,
\end{equation}
and satisfying the following system of ODEs
\begin{equation}
\label{eq:system 3}
 \left\{
\begin{array}{l}
-A_{1i}'(s)-2\lambda_{\theta}A_{1i}(s)+\sigma_{\theta}^2 A_{1i}(s)^2+\frac{\g_1}{(1-\g_1)^2}=0\\
-A_{2i}'(s)+\lambda_{\theta}(\bar{\theta}A_{1i}(s)-A_{2i}(s))+\sigma_{\theta}^2A_{1i}(s)A_{2i}(s)=0\\
-A_{3i}'(s)+\lambda_{\theta}\bar{\theta}A_{2i}(s)+\frac{\sigma_{\theta}^2}{2}(A_{1i}(s)+A_{2i}(s)^2)=0
\end{array}
\right. \;.
\end{equation}
 Define for $i=1,2,3 $ the function $H_i:[0,T] \times \R\to (0,\infty)$ by
 $$
H_i(\tau, \theta):= \exp\left(A_{1i}(\tau)\frac{\theta^2}{2}+A_{2i}(\tau)
\theta+A_{3i}(\tau)-\frac{r(1-\g_i)+\rho_i}{1-\g_i}\tau\right)
\;,
$$
\end{definition}
The following holds:
\begin{proposition}
 \label{prop:semi-explicit portfolio process}
Define the process  $Y_t, t\in[0,T]$ by \begin{equation}
Y_t:=(yZ_t)^{-1}e^{rt},
\end{equation}
with $y=\Y(x)$ and the process $Z_t, t\in[0,T]$ given by Definition \eqref{Zt}. 
{
For $i=1, 2, 3$, we have:
 \begin{equation}
 \Xeye_t = F^i(t, \theta_t, Y_t)
 \end{equation}
 where
\begin{equation}
F^i(t, \theta, y) = \begin{cases}
y^{\frac{1}{1-\g_i}}e^{\frac{(r+\rho_i)t}{1-\gamma_i}}\int_0^{T-t} H_i(\theta, \tau) d\tau \textrm{ for } i=1,2\\
y^{\frac{1}{1-\g_3}}e^{\frac{(r+\rho_3)t}{1-\gamma_3}} H_3(\theta, T-t) \textrm{ for } i=3
\end{cases}
\end{equation}
}
\end{proposition}

\proof
Following Theorem \ref{theorem},
the optimal initial allocation is given by $x_i=\H_i(\Y(x))$.
Denoting $y:=\Y(x)=\Y_i(x_i)$, 
the theorem gives also the optimal
consumption

$$
\Ceye_t(x_i) = I_i(y\zeta_t^i)=(y \exp(\rho_i t ) Z_t)^{\frac{1}{\g_i -1}}
\; , \quad 0\leq t\leq T \;,
$$
where $Z_t, t\in[0,T]$ is the state density process defined by \eqref{Zt}.
By Ito's formula, \begin{equation}
dY_t = (r+\theta_t^2)Y_tdt +\theta_t Y_t dW_t
\end{equation}
The optimal total wealth process of agent $i$ is thus given by

$$
\Xeye_t=\tilde{\E}\left[\int_t^T D_s \Ceye_s ds |\F_t\right], t\in[0,T].
$$
We have the relations $\Ceye_s=(ye^{\rho_i s }Z_s)^{\frac{1}{\g_i-1}}=(e^{\rho_i s }e^{rs}Y_s^{-1})^{\frac{1}{\g_i-1}}$ ,  $\frac{Z_s}{Z_t}=e^{r(s-t)}\frac{Y_t}{Y_s}$ and $D_s=e^{-rs}$.

For $i=1, 2, 3$, define $f^i(t,\tau,\theta)$  by:
$$
f^i(t,\tau,\theta) =\E\left[\exp\left(\frac{\g_i}{2(1-\g_i)}\int_t^{\tau} 
\theta_s^2 ds+\frac{\g_i}{1-\g_i}\int_t^{\tau}  \theta_s dW_s  \right)  | \theta_t=\theta\right].
$$	So, for $i=1, 2$
{ \begin{eqnarray*}
	\Xeye_t  & =& \E\left[\int_t^T \frac{Z_s}{Z_t}D_s \Ceye_s ds |\F_t\right] \\
 &=&	 \E_t\left[\int_t^T e^{r(s-t)} \frac{Y_t}{Y_s}e^{-r s} Y_s^{\frac{1}{1-\gamma_i}} e^{(r+\rho_i)\frac{s}{\gamma_i-1}}ds \right]\\
	&=& Y_t^{1+\frac{\g_i}{1-\g_i}}\E_t\left[  \int_t^T e^{-r t  -\frac{(\rho_i+r) s}{1-\g_i} } \left( \frac{Y_s}{Y_t} \right)^{\frac{\g_i}{1-\g_i}} ds \right] \\ 
		&=& Y_t^{\frac{1}{1-\g_i}}  \int_t^T e^{-r t  -\frac{(\rho_i+r) s}{1-\g_i}+\frac{r\g_i (s-t)}{1-\g_i}}f^i(t,s,\theta_t)ds 
\end{eqnarray*}
It is easy to see that we just have to replace $s$ by $T$ in the integral term to obtain $\Xthree_t$:
\begin{eqnarray*}
	\Xthree_t	&=& Y_t^{\frac{1}{1-\gamma_3}} e^{-r t  -\frac{(\rho_3+r) T}{1-\g_3}+\frac{r\g_3 (T-t)}{1-\g_3}}  f^3(t,T,\theta_t)
\end{eqnarray*}
}
For, $0\leq t  \leq \tau$, the process 

$$
f^i(t,\tau,\theta_t)\exp\left(\frac{\g_i}{2(1-\g_i)}
\int_0^t  \theta_s^2 ds+\frac{\g_i}{1-\g_i}\int_0^t  \theta_s dW_s  \right) \; ,
$$
is a conditional expectation of a $\Fc_T$-measurable random variable for any fixed $\theta$.
It is then a $\P$-martingale on time $\tau\leq T$. 
Notice that in the definition of $f^i(t,\tau,\theta),$ the coefficients of the exponential are independent of $t$.
Therefore we look for $f^i(t,\tau,\theta)$ of the form $f^i(t,\tau,\theta_t):=g^i(\tau-t,\theta_t)$.
We make the change of variables $\tau-t:=s$. 
Given that the function $(s,\theta)\mapsto g^i (s,\theta)$ is $\Cc^{1,2},$
it follows by Ito's formula that
\begin{equation}\label{PDE}
-g^i_s-\lambda_{\theta}(\theta-\bar{\theta})g^i_{\theta}
+\frac{\sigma_{\theta}^2}{2}g^i_{\theta\theta}+\left(\frac{\g_i}{2(1-\g_i)}
+\frac{\g_i^2}{2(1-\g_i)^2}\right)\theta^2 g^i=0 \; .
\end{equation}
with the condition $g^i(0,\theta)=1$. 
We follow \cite{Wachter} and search for $g^i(s,\theta)$ of the form

$$
g^i(s,\theta)=\exp\left(A_{1i}(s)\frac{\theta^2}{2}+A_{2i}(s)\theta+A_{3i}(s))\right)\; ,
$$ 
where $A_{1i}, A_{2i}, A_{3i}$ are three continuous functions of $s$.
The terminal condition in the latter expression implies condition \eqref{eq:boundary condition of A}.
Plugging the expression of $g^i(s,\theta)$ in \eqref{PDE}, we get a second-order polynomial in $\theta$
\begin{eqnarray*}
&& \theta^2\left[ -A_{1i}'(s)-2\lambda_{\theta}A_{1i}(s)+\sigma_{\theta}^2 A_{1i}(s)^2+\frac{\g_i}{(1-\g_i)^2}\right]\\
&& +\theta \left[ -A_{2i}'(s)+\lambda_{\theta}(\bar{\theta}A_{1i}(s)-A_{2i}(s))+\sigma_{\theta}^2A_{1i}(s)A_{2i}(s) \right]\\ 
&& -A_{3i}'(s)+\lambda_{\theta}\bar{\theta}A_{2i}(s)+\frac{\sigma_{\theta}^2}{2}(A_{1i}(s)+A_{2i}(s)^2) = 0 \; 
\end{eqnarray*}

Since the equation holds for any $\theta\in \R$,
we separate the coefficients in $\theta^2, \theta$ and constant to obtain \eqref{eq:system 3}.
Thus, the functions $g^i$ and $H_i$ are equal.
 \ep

 If a $C^1$ function $s\mapsto A_{1i}(s), s\in [0,T-t]$ has been found,
then $s\mapsto A_{2i}(s)$ is given by a linear ODE,
which finally allows to retrieve $s\mapsto A_{3i}(s):$

$$
A_{3i}(s)=\int_0^{s} \lambda_{\theta}\bar{\theta}A_{2i}(u)+\frac{\sigma_{\theta}^2}{2}(A_{1i}(u)+A_{2i}(u)^2)du \; .
$$
This allows for the detailed numerical analysis we present in subsection \ref{sec:numerics}.
The following provides the missing part.

\begin{proposition}
 \label{prop:semi-explicit solution 2}
 Let $A_{1i}(s)$ be a solution on $\{0\le t \le T-t\}$ of the ODE
 \begin{equation}
  \label{eq:ODE}
   -A_{1i}'(s)-2\lambda_{\theta}A_{1i}(s)+\sigma_{\theta}^2 A_{1i}(s)^2+\frac{\g_i}{(1-\g_i)^2}=0
 \end{equation}
 such that $A_{1i}(0)=0$. Then, denoting $\Delta=\lambda_{\theta}^2-\frac{\g_i \sigma_{\theta}^2}{(1-\g_i)^2}$,
the function $s\mapsto A_{1i}(s)$ is defined on $[0,T-t]$ by
 \begin{equation}
  \label{eq:solution of A_1}
  A_{1i}(s)= \left\{
  \begin{array}{lll}
\dfrac{\g_i}{(1-\g_i)^2\sigma_{\theta}}\dfrac{1-\exp(-2\sqrt{\Delta}\sigma_{\theta}^2s)}{\lambda_{\theta}+\sqrt{\Delta}-(\lambda_{\theta}-\sqrt{\Delta})\exp(-2\sqrt{\Delta}\sigma_{\theta}^2s)}\\
  & \text{if }\Delta>0 \\
 \dfrac{\lambda_{\theta}^2 s}{\sigma_{\theta}^2\lambda_{\theta}s +\sigma_{\theta}^2} 
 & \text{if }\Delta=0 \\
  \dfrac{1}{\sigma_{\theta}^2}\left(\sqrt{-\Delta}\tan\left(-\sqrt{-\Delta} s+\arctan\left(\frac{-\lambda_\theta}{\sqrt{-\Delta}}\right) \right) 
  +\lambda_{\theta}\right)
  & \text{if }\Delta<0
  \end{array}
  \right. \; .
 \end{equation}
\end{proposition}

\proof 

\textit{Case 1: $\Delta>0$.}
There are two distinct roots 
to the characteristic polynomial of the ODE, 
given by
$m_{\pm}:=\frac{\lambda_{\theta}}{\sigma_{\theta}}\pm \sqrt{\Delta}$.
A general solution $A_1(t)$ to \eqref{eq:ODE} shall verify
\begin{eqnarray*}
	\sigma_{\theta} &=& \frac{\sigma_{\theta}A_{1i}'(\tau)}{(\sigma_{\theta}A_{1i}(\tau)-\frac{\lambda_{\theta}}{\sigma_{\theta}})^2-\sigma_{\theta}^2\Delta}\\
	\sigma_{\theta}s &=& \int_0^s \frac{\sigma_{\theta}A_{1i}'(\tau)d\tau}{(\sigma_{\theta}A_{1i}(\tau)-\frac{\lambda_{\theta}}{\sigma_{\theta}}-\sigma_{\theta}\sqrt{\Delta})(\sigma_{\theta}A_{1i}(\tau)-\frac{\lambda_{\theta}}{\sigma_{\theta}}+\sigma_{\theta}\sqrt{\Delta})}\\
	&=& \int_0^{A_{1i}(s)} \frac{\sigma_{\theta}dx}{(\sigma_{\theta}x-\sigma_{\theta}m_{+})(\sigma_{\theta}x-\sigma_{\theta}m_{-})}\\
	&=&  \frac{1}{\sigma_{\theta} (m_{+}-m_{-})}\int_0^{A_{1i}(s)} \frac{dx}{x-m_{+}}-  \frac{dx}{x-m_{-}}\;, \\
	&=&  \frac{1}{2\sigma_{\theta} \sqrt{\Delta}}\log \left(\frac{(A_{1i}(s)-m_{+})m_{-}}{(A_{1i}(s)-m_{-})m_{+}}\right)\;,
\end{eqnarray*}
Therefore,

$$
\frac{A_{1i}(s)-m_{+}}{A_{1i}(s)-m_{-}}=\frac{m_{+}}{m_{-}}\exp\left(2\sqrt{\Delta}\sigma_{\theta}^2s\right)
$$
with 

$$
	A_{1i}(s)= m_+ +\frac{2\sigma_{\theta}\sqrt{\Delta}}{1-\frac{m_{+}}{m_{-}}\exp\left(2\sqrt{\Delta}\sigma_{\theta}^2s\right)}
	\quad \forall t\le \tau \;.
$$

\textit{Case 2: $\Delta=0$.}
With the double root $m:=\lambda_{\theta}/\sigma_{\theta}^2$,
the same operation provides
\begin{eqnarray*}
	\sigma_{\theta} &=& \frac{A_{1i}'(\tau)}{(A_{1i}(\tau)-m)^2}\\
	\sigma_{\theta}s &=&\int_0^s\frac{A_{1i}'(\tau)}{(A_{1i}(\tau)-m)^2}d\tau\\
	\sigma_{\theta}s&=& \left(-\frac{1}{m}+\frac{1}{m-A_{1i}(s)}\right)\; .
\end{eqnarray*} 
The solution then follows:

\begin{eqnarray*}
	A_{1i}(s) &=& \frac{\sigma_{\theta}m^2s}{1+m\sigma_{\theta}s}
\end{eqnarray*} 

\textit{Case 3: $\Delta<0$.}
We can write the ODE as
$$
-A_{1i}'(s)+\left(\sigma_{\theta}A_{1i}(s)-\frac{\lambda_{\theta}}{\sigma_{\theta}}\right)^2-\frac{\Delta}{\sigma_{\theta}^2}=0 \; .
$$
Taking 
$y(s):=(\sigma_{\theta}^2 A_{1i}(s) -\lambda_{\theta})/\sqrt{-\Delta'}$, 
we get 
$$
\arctan (y(s))-\arctan\left(-\frac{\lambda_{\theta}}{\sqrt{-\Delta'}}\right)  = -\sqrt{-\Delta} s
$$
providing the solution.
\ep

\vspace{0.4cm}
Notice that $s\mapsto A_{1i}(s)$ is not continuous nor well defined for all $s$, if $\Delta\le 0$.
The condition $\Delta>0$  can also write
$$
\g_1 <\g_{lim}:=\frac{2}{2+b+\sqrt{(2+b)^2-4}} \text{ for } b:=\frac{\sigma_{\theta}^2}{\lambda_{\theta}^2}\;.
$$
Proposition \ref{prop:semi-explicit portfolio process} provides the portfolio process value for a consuming agent.

{
\begin{corollary}
Let $s_i = \int_0^T H_i(\theta_0, s)ds$, for $i=1, 2$ and $s_3=H_3(\theta_0, T)$. 
 The initial allocations for the three agents are:
\begin{equation}
\label{ei}
\left\{
\begin{array}{lll}
\displaystyle x_1 &=&  y^{\frac{1}{\g_1-1}}\int_0^T H_1(s,\theta_0)ds =  y^{\frac{1}{\g_1-1}} s_1\\
\displaystyle x_2 &=&  y^{\frac{1}{\g_2-1}}\int_0^T H_2(s,\theta_0)ds = y^{\frac{1}{\g_2-1}} s_2\\
x_3 &=&  y^{\frac{1}{\g_3-1}} H_3(T,\theta_0) = y^{\frac{1}{\g_3-1}}s_3
\end{array}\right.
\end{equation}
where $y=\Y(x)$ is uniquely defined such that $x_1+x_2+x_3=x$.
\end{corollary}
}

{
\begin{proposition}
The investment $\pieye$ corresponding to agent $i$ is given by
\begin{equation}
\pieye_t = \frac{1}{\sigma F^i}\big(Y_t\theta_t \frac{\partial F^i}{\partial Y}-\sigma_{\theta}\frac{\partial F^i}{\partial \theta}(t,\theta_t,Y_t)\big)
\end{equation}
\end{proposition}
\proof
We apply It\^o's lemma to the equality $\Xeye_t=F^i(t,\theta_t,Y_t)$.
\begin{eqnarray*}
d\Xeye_t&=&(r \Xeye_t -\Ceye_t)dt +b(t)\pieye_t dt+\sigma \pieye_t dW_t= (\mathcal{L} F^i+\frac{\partial F^i}{\partial t}) \; dt+(\frac{\partial F^i}{\partial \theta}(-\sigma_{\theta})+\frac{\partial F^i}{\partial Y}Y_t\theta_t) dW_t \label{dFt}.
\end{eqnarray*}
Writing the equality between the  $dW_t$ terms yields:
\begin{eqnarray*}
\sigma \pieye_t F^i &=& Y\theta \frac{\partial F^i}{\partial Y}-\sigma_{\theta}\frac{\partial F^i}{\partial \theta}\label{portfolioi}\\
\end{eqnarray*}
which ends the proof.
\ep
}

The economic consequences of these equations are explored in the next subsection.
We continue here to explore the analytical results.
\begin{proposition}\label{MPRdependence}
Assume that $\theta_t>0$ (positive MPR). If $\gamma_i<0,$ $i=1, 2,$ the CSP decreases with $\theta$.  
On the other hand if $\gamma_i\geq0,$ $i=1, 2,$ the CSP increases with $\theta.$
\end{proposition}

\proof
From direct computations one gets 
$$
\frac{dy}{d\theta}
=\frac{y^{\frac{1}{\g_1-1}}ds_1 + y^{\frac{1}{\g_2-1}}ds_2}
{\left(\frac{s_1}{1-\g_1}y^{\frac{1}{\g_1-1}-1}+\frac{s_2}{1-\g_2}y^{\frac{1}{\g_2-1}-1}+\frac{s_3}{1-\g_3}y^{\frac{1}{\g_3-1}-1}\right)d\theta} \; .
$$
Moreover
$$
\frac{d}{d\theta}\left(\frac{x_1 + x_2 }{x}\right)=\frac{d}{d\theta}\left(1-\frac{x_3}{x}\right)=-\frac{dx_3}{xd\theta}
=\frac{y^{\frac{1}{\g_3-1}-1}dy}{(1-\g_3)xd\theta}.
$$
If $\theta_t>0$ and $\g_i<0$, it follows from the monotonicity of $A_{1i}$ and $A_{2i}$ that
$$
A_{1i}(0,T-t)\theta_t+A_{2i}(0,T-t)\leq\int_0^{T-t}p_i(\theta_t,t,\tau)(A_{1i}(0,\tau)\theta_t+A_{2i}(0,\tau))d\tau\leq 0\;.
$$
On the other hand if $\g_i\geq 0,$
$$
0\leq\int_0^{T-t}p_i(\theta_t,t,\tau)(A_{1i}(0,\tau)\theta_t+A_{2i}(0,\tau))d\tau\leq A_{1i}(0,T-t)\theta_t+A_{2i}(0,T-t)\;.
$$
\ep

During favorable market conditions, i.e., when $\theta$ is increasing, the agents behave differently according to their risk aversion. 
Thus, if they are more risk averse they will use a higher fraction of the initial wealth to finance investment; 
else if they are less risk averse they will use a higher fraction of the initial wealth to finance consumption.

\begin{proposition}
Assume that $\g_1<\g_2<\g_3.$
Then we get the following assymptotics for the couple risk aversion 
$$
\lim_{x\rightarrow \infty} R(x)=(1-\g_1)\frac{\H_3(1)}{\H_1(1)} \quad \text{and}\quad 
\lim_{x\rightarrow 0} R(x)=(1-\g_3)\frac{\H_1(1)}{\H_3(1)}\; .
$$
\end{proposition}

\proof
Recall that 
$V(x)=G(\Y(x))$ and 
$V'(x)=\Y(x)$.
Thus $V''(x)=\Y'(x)=1/\H'(y)$ (with $y:=\Y(x)$) and 
$R(x)=-x\Y'(x)/\Y(x)=-\H(y)/(y\H'(y))$. 
In light of  
$$
\H(y)=y^{\frac{1}{\g_1-1}}\H_1(1)+y^{\frac{1}{\g_2-1}}\H_2(1)+y^{\frac{1}{\g_3-1}}\H_3(1) \;,
$$ 
it follows that
$$
R(x)=\frac{y^{\frac{1}{\g_1-1}}\H_1(1)+y^{\frac{1}{\g_2-1}}\H_2(1)+y^{\frac{1}{\g_3-1}}\H_3(1)}{y^{\frac{1}{\g_1-1}}
\frac{\H_1(1)}{1-\g_1}+y^{\frac{1}{\g_2-1}}\frac{\H_2(1)}{1-\g_2}+y^{\frac{1}{\g_3-1}}\frac{\H_3(1)}{1-\g_3}}\;,
$$
whence the claim.
\ep

For small initial wealth or high initial wealth
the couple risk aversion is driven by one of the agents. 
Thus, the less risk averse agent
determines the couple's utility for little initial wealth. 
This is in accordance with risk seeking agents
behavior when the latter are poor.

\subsection{Numerical results}
\begin{figure}[ht]
\centering
\includegraphics[scale=0.7]{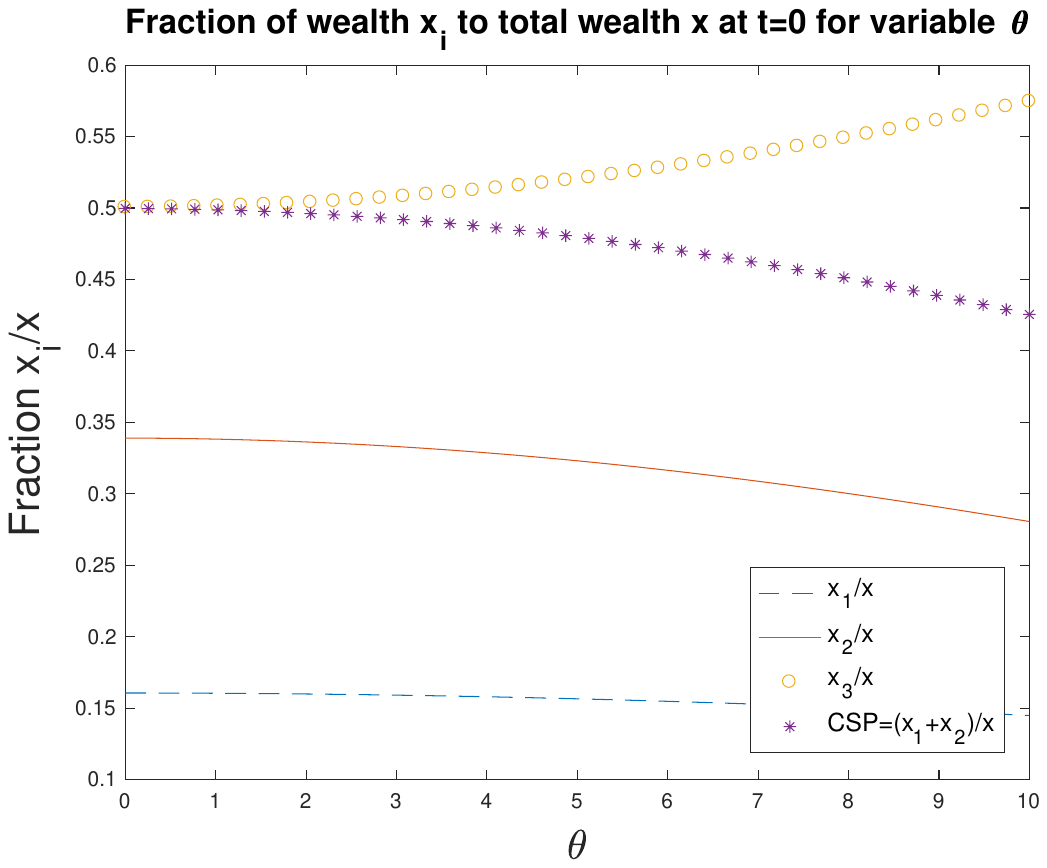}
\vspace{-2cm}
\caption{Fraction of initial wealth as a function of $\theta$.}
\label{fig:1}
\end{figure}

In the section \ref{Del}  the consumption satisfaction proportion (CSP) dependence on the initial wealth was explored.
In this section we further study CSP dependence on other model parameters. More precisely, placing ourselves in the
setting of section \ref{meanreverting}, and taking advantage of the closed form formulas established there,
we investigate the CSP dependence with respect to market price of risk in one hand and the agents risk aversion
on the other hand. Our closed formulas are explicit up to computing some integrals, task which we perform using
the modified Euler numerical scheme. In our numerical experiments we have chosen the following financial market parameters:

$$
(r,\sigma_{\theta}, \lambda_{\theta}, \bar{\theta}) = (0.048, 0.0655, 0.2712, 0.9456) \; .
$$
The (total) initial wealth $x$ is set to $10,$ i.e., $x=10$. In figure \ref{fig:1} we observe the effect on CSP of the
market price of risk $\theta.$ Here $(\g_1, \g_2, \g_3)=(-9, -3, -2).$
The findings are in accordance with Proposition \ref{MPRdependence}. The agents will use a higher fraction of the initial wealth to finance investment in favorably financial market conditions since they are risk averse. This in turn will make CSP decrease with respect to the market price of risk. By looking at figure \ref{fig:1} we notice that the effect of market price of risk on CSP is nearly marginal; a change
in market price of risk will cause at most $8\%$ change in CSP.

\begin{figure}[ht]
\centering
\includegraphics[scale=0.7]{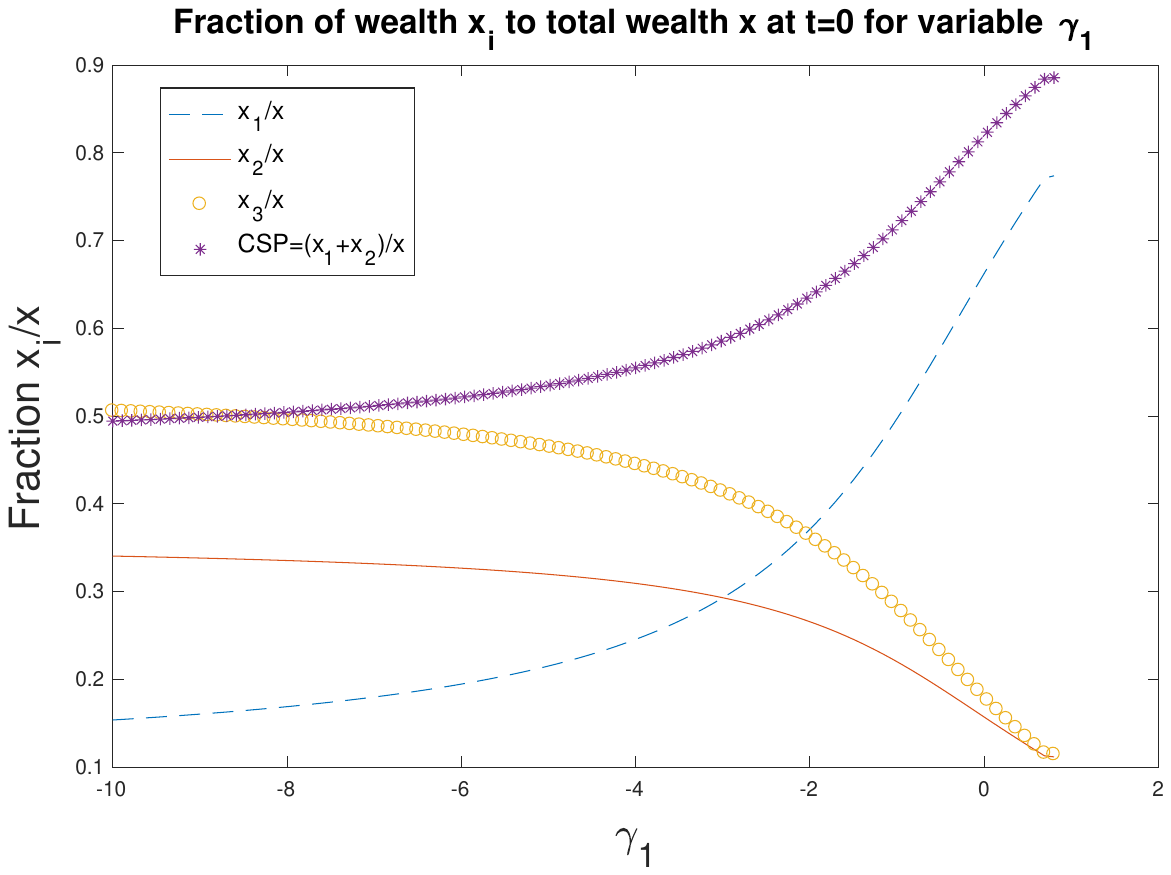}
\vspace{-4cm}
\caption{Fraction of initial wealth as a function of risk aversion $\g_1$, for $\g_1\le \g_3$.}
\label{fig:2}
\end{figure}

Let us next explore the effect of first agent change in risk aversion on CSP. In figure \ref{fig:2},
we vary $\g_1$ while holding $\g_2=-3,$ and $\g_3=-2.$ As expected, when agent 1 becomes more risk-averse 
his initial wealth allocation for financing his/her consumption increases and this will translate in an increase in CSP.
By looking at figure \ref{fig:2} we notice that the effect of first agent change in risk aversion on CSP is considerable; a change
in first agent risk aversion will cause up to $50\%$ change in CSP.

\begin{figure}[ht]
\centering
\includegraphics[scale=0.7]{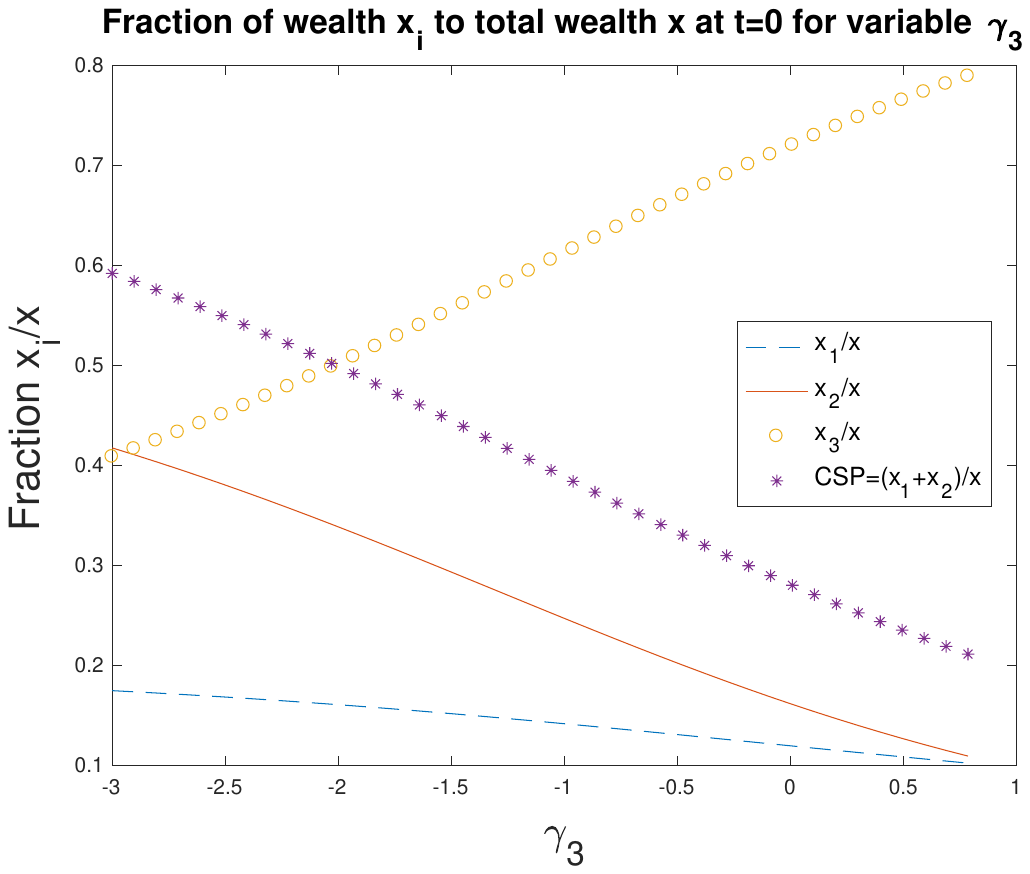}
\vspace{-4cm}
\caption{Fraction of initial wealth as a function of $\g_3$.}
\label{fig:3}
\end{figure}

We conclude by exploring the effect of the couple change in risk aversion on CSP. In figure \ref{fig:3},
we vary $\g_3$ while holding $\g_1=-9,$ and $\g_2=-3.$ As expected the initial wealth allocation for financing investment increases in $\g_3$
(since the couple becomes less risk averse). This in turn will make CSP decrease in $\g_3.$ By looking at figure \ref{fig:3} we notice that the effect of the couple change in risk aversion on CSP is considerably; a change
in couple risk aversion will cause up to $70\%$ change in CSP.

\bibliographystyle{plain}
\bibliography{myrefs}

\begin{thebibliography}{10}

\bibitem{browning2000saving}
Martin Browning.
\newblock The saving behaviour of a two-person household.
\newblock {\em scandinavian Journal of Economics}, 102(2):235--251, 2000.

\bibitem{browning1996household}
Martin Browning and Annamaria Lusardi.
\newblock Household saving: Micro theories and micro facts.
\newblock {\em Journal of Economic literature}, 34(4):1797--1855, 1996.

\bibitem{CH}
P~Cheridito and Y.~Hu.
\newblock Optimal consumption and investment in incomplete markets with general
  constraints.
\newblock {\em Stochastics and Dynamics}, 11(2):283--299, 2011.

\bibitem{chiappori1988rational}
Pierre-Andr{\'e} Chiappori.
\newblock Rational household labor supply.
\newblock {\em Econometrica: Journal of the Econometric Society}, pages 63--90,
  1988.

\bibitem{chiappori1992collective}
Pierre-Andre Chiappori.
\newblock Collective labor supply and welfare.
\newblock {\em Journal of political Economy}, 100(3):437--467, 1992.

\bibitem{chiappori2017static}
Pierre-Andre Chiappori and Maurizio Mazzocco.
\newblock Static and intertemporal household decisions.
\newblock {\em Journal of Economic Literature}, 55(3):985--1045, 2017.

\bibitem{ekeland2012time}
Ivar Ekeland, Oumar Mbodji, and Traian~A Pirvu.
\newblock Time-consistent portfolio management.
\newblock {\em SIAM Journal on Financial Mathematics}, 3(1):1--32, 2012.

\bibitem{Fama1}
E.~Fama and K.~French.
\newblock Permanent and temporary components of stock prices.
\newblock {\em Journal of Political Economy 96, 246 - 273}, 1988.

\bibitem{harrison1981martingales}
J.M. Harrison and S.R. Pliska.
\newblock Martingales and stochastic integrals in the theory of continuous
  trading.
\newblock {\em Stochastic processes and their applications}, 11(3):215--260,
  1981.

\bibitem{HRIH}
H~Horst, Y.~Hu, Imkeller, A~Reveillac, and J.~Zhang.
\newblock Forward-backward systems for expected utility maximization.
\newblock {\em Stochastic Processes and their Applications}, to appear, 2014.

\bibitem{HIM}
Y.~Hu, P~Imkeller, and M.~M\"uller.
\newblock Utility maximization in incomplete markets.
\newblock {\em Ann. Appl. Probab.}, 15(3):1691--1712, 2005.

\bibitem{HuangLitzenberger}
C.~Huang and R.~H. Litzenberger.
\newblock Foundations for financial economics.
\newblock 1988.

\bibitem{KaratzasSmallInvestor}
I.~Karatzas, J.~P. Lehoczky, and S.~E. Shreve.
\newblock Optimal portfolio and consumption decisions for a small investor; on
  a finite horizon.
\newblock {\em SIAM J. Control Optim.}, 25(6):1557--1586, November 1987.

\bibitem{KaratzasShreve1987}
I.~Karatzas and S.~E. Shreve.
\newblock Brownian motion and stochastic calculus.
\newblock {\em Graduate Texts in Mathematics}, 113, 1987.

\bibitem{Kim}
T.S. Kim and E.~Omberg.
\newblock Dynamic nonmyopic portfolio behavior.
\newblock {\em Review of Financial Studies}, 9(1):141--161, 1996.

\bibitem{manser1980marriage}
Marilyn Manser and Murray Brown.
\newblock Marriage and household decision-making: A bargaining analysis.
\newblock {\em International economic review}, pages 31--44, 1980.

\bibitem{mazzocco2007household}
Maurizio Mazzocco.
\newblock Household intertemporal behaviour: A collective characterization and
  a test of commitment.
\newblock {\em The Review of Economic Studies}, 74(3):857--895, 2007.

\bibitem{meyer2005}
D~Meyer and J.~Meyer.
\newblock Relative risk aversion: What do we know?
\newblock {\em Journal of Risk andUncertainty}, 31(3):243--262, 2005.

\bibitem{ortigueira2013important}
Salvador Ortigueira and Nawid Siassi.
\newblock How important is intra-household risk sharing for savings and labor
  supply?
\newblock {\em Journal of Monetary Economics}, 60(6):650--666, 2013.

\bibitem{osfc}
OSFC.
\newblock Evidence for mean reversion in equity prices.
\newblock {\em Office of the Superintendent of Financial Institutions of
  Canada}, March, 2012.

\bibitem{Summers}
J.~Poterba and L.H. Summers.
\newblock Mean reversion in stock prices: Evidence and implications.
\newblock {\em Journal of Financial Economics 22, 27 - 59}, 1988.

\bibitem{Six_optimalconsumption}
P.~Six.
\newblock Dynamic strategies when consumption and wealth risk aversion differ.
\newblock {\em Revue Finance}, 31:93--118, 2010.

\bibitem{Wachter}
J.A. Wachter.
\newblock Portfolio and consumption decisions under mean-reverting returns: An
  exact solution for complete markets.
\newblock {\em Journal of financial and quantitative analysis}, 2002.

\bibitem{WH}
B~Wong and C.~C. Heyde.
\newblock On changes of measure in stochastic volatility models.
\newblock {\em Journal of Applied Mathematics and Stochastic Analysis}, pages
  1--13, 2006.

\end{thebibliography}

\end{document}